\documentclass[aps,prl,twocolumn,showpacs,superscriptaddress]{revtex4-1}

\usepackage[final]{pdfpages}

\usepackage{soul}
\usepackage{mathrsfs}
\usepackage{amsfonts}
\usepackage{amssymb}
\usepackage{amsmath}
\usepackage{graphicx}
\usepackage{sidecap}
\usepackage{color}
\usepackage{subeqnarray}
\usepackage{verbatim}
\usepackage{ulem}
\usepackage{subfigure}
\usepackage{graphicx}
\usepackage{braket}
\usepackage{amsmath}
\usepackage[T1]{fontenc}
\usepackage{textcomp}
\usepackage{gensymb}

\def\bea{\begin{eqnarray}}
\def\eea{\end{eqnarray}}

\def\B{\mathcal{B}}
\def\H{\mathcal{H}}
\def\W{\widetilde{W}}
\def\I{\mathrm{I}}

\def\tr{\mathrm{tr}}

\begin{document}


\title{Entanglement Detection with Single Hong-Ou-Mandel Interferometry}

\author{ Chang Jian Kwong}
\affiliation{Centre for Quantum Technologies, National University of Singapore, 3 Science Drive 2, Singapore 117543, Singapore}
\author{ Simone Felicetti}
\affiliation{ Department of Physical Chemistry, University of the Basque Country UPV/EHU, Apartado 644, E-48080 Bilbao, Spain}
\affiliation{ Laboratoire Mat\' eriaux et Ph\' enom\`enes Quantiques, Sorbonne Paris Cit\' e, Universit\' e Paris Diderot, CNRS UMR 7162, 75013, Paris, France}
\author{Leong Chuan Kwek}
\affiliation{Centre for Quantum Technologies, National University of Singapore, 3 Science Drive 2, Singapore 117543, Singapore}
\affiliation{Institute of Advanced Studies, Nanyang Technological University, 60 Nanyang View, Singapore 639673, Singapore}
\affiliation{National Institute of Education, Nanyang Technological University, 1 Nanyang Walk, Singapore 637616, Singapore}
\affiliation{MajuLab, CNRS-UNS-NUS-NTU International Joint Research Unit, UMI 3654, 117543, Singapore }
\author{ Joonwoo Bae}
\email{joonwoobae@hanyang.ac.kr}
\affiliation{ Department of Applied Mathematics, Hanyang University (ERICA), 55 Hanyangdaehak-ro, Ansan, Gyeonggi-do, 426-791, Korea}


\begin{abstract}
Entanglement is not only fundamental for understanding multipartite quantum systems but also generally useful for quantum information applications. Despite much effort devoted so far, little is known about minimal resources for detecting entanglement and also comparisons to tomography which reveals the full characterization. Here, we show that all entangled states can in general be detected in an experimental scheme that estimates the fidelity of two pure quantum states. An experimental proposal is presented with a single Hong-Ou-Mandel interferometry in which only two detectors are applied regardless of the dimensions or the number of modes of quantum systems. This shows measurement settings for entanglement detection are in general inequivalent to tomography: the number of detectors in quantum tomography increases with the dimensions and the modes whereas it is not the case in estimation of fidelity which detects entangled states. 
\end{abstract}


\maketitle


Entanglement is generally a useful resource in quantum information processing \cite{ref:reviewh} and it provides a deeper understanding of quantum many-body systems \cite{ref:recent} \cite{ref:nature}. The problem of determining whether given states are entangled or separable, nevertheless, turns out to be hard in general \cite{ref:separability}. It is also non-trivial to detect the presence of entanglement before the full characterization of quantum states such as tomography. 

The most general and feasible method of detecting entangled states \cite{ref:reviewew, ref:entdec} involves the construction of entanglement witnesses (EWs) \cite{ref:ewterhal}. These are essentially observables described by positive-operator-valued-measures (POVMs) that correspond to physical detectors in experiment. Since entanglement is just one facet in the characterization of multipartite quantum systems, entanglement detection would correspond to merely a part of the tomography needed for the full characterization of quantum states. In fact, EWs form an incomplete measurement, a subset of the required measurements needed for quantum state tomography where a complete measure is performed, see e.g. \cite{ref:m}. For instance, two local measurement settings over given systems suffice to construct EWs that detect genuine multipartite entanglement \cite{ref:gt}.







On the other hand, quantum state tomography provides the most comprehensive scheme for detecting entangled states. Once a complete experimental characterization of a the state of a given system is obtained, numerical or analytical techniques are employed to decide if the state of the system is entangled or separable, see e.g. \cite{ref:reviewh}. It turns out that if single-copy observables such as sets of EWs are applied to completely determining whether quantum states are entangled or separable, POVMs of the observables can actually be used to perform tomography \cite{ref:laflamme}. That is, to completely determine whether givens states are entangled or separable, one naturally resorts to tomography for entanglement detection. A set of EWs may correspond to a partial tomography such that if the set has a sufficient number of EWs, their POVMs can actually be used to perform quantum state tomography.



In this work, we approach to entanglement detection with minimal resources and show that measurement of EWs does not generally correspond to a partial tomography. Namely, the number of detectors increases in tomography for larger and higher-dimensional systems, but it appears that implementation of EWs may not the case in general. We show that any experimental scheme to estimate fidelity of two pure states, such as the Hong-Ou-Mandel (HOM) interferometry for two single-photon sources, can detect any type of entangled states. In particular, we present an experimental proposal based on a single HOM interferometry where a measurement setting contains only two detectors \cite{ref:HOM}. The detection method is feasible with current experimental technologies. Our scheme relies on two recent techniques in photonics, quantum state joining \cite{ref:quantumjoin} and the embedding of orbital-angular-momentum degrees (OAM) of freedom of single photons \cite{ref:eOAM}, prior to the measurement. In contrast to quantum tomography, only two detectors of the HOM interferometer are needed in the measurement regardless of dimensions and modes of given systems. The detection scheme is valid to all physical bosonic system where the particles exhibit bunching. 

Let us begin by introducing the problem of detecting entangled states with EWs. We denote  $\B(\H)$ as the set of bound linear operators in a Hilbert space $\H$ and with $S(\H)$ as the set of quantum states in $\H$. EWs correspond to Hermitian operators, i.e. observables, $\{ W^{\dagger} = W \in \B(\H\otimes \H)\}$ such that 
\bea
\tr[\sigma_{\mathrm{sep}} W] \geq 0 &~~& \mathrm{for~all~separable~states}~\sigma_{\mathrm{sep}},~\mathrm{and}  \nonumber \\
\tr[\rho_{\mathrm{ent}} W] < 0 &~~& \mathrm{for~some~entangled~states}~\rho_{\mathrm{ent}}. \label{eq:aewdec}  
\eea 
One can restrict EWs to cases of unit trace, i.e. $\tr[W]=1$. Negative expectation values of EWs unambiguously lead to the conclusion that the given unknown states are entangled. Experimental implementation of EWs is feasible with current technologies and works as follows. First, an EW is decomposed into a linear combination of POVMs, i.e. $W = \sum_{i} c_i M_i$ with POVMs $\{ M_i \geq 0\}$ for some constants $\{ c_i\}$ \cite{ref:loccEW}. For a particular decomposition of an EW, each POVM element $M_i$ describes a measurement detector. The probability of detecting an event at $M_i$ is given by $p(M_i) =\tr[\rho M_i] $ for a state $\rho$. In this way, probabilities $p(M_i)$ are obtained in an experiment for all $i$ and, the expectation $\tr[\rho W]$ is then obtained by computing $\sum_i c_i p(M_i)$ with constants $\{ c_i\}$. If the expectation value is negative, the state is determined to be entangled. Note that if each $M_i$ can be prepared by local operations and classical communication (LOCC), then the detection scheme can be implemented by LOCC \cite{ref:loccEW}. This is particularly useful for cases where quantum states are shared by remotely separated parties. 

We now map EWs into nonnegative operators called approximate EWs (AEWs). Given a witness $W$, the mapping to a non-negative operator $\W$ is given by
\bea 
\W = (1-p^{*})W + p^{*} \I  \label{eq:aewcon}
\eea 
with the minimal $p^{*}$ such that $\W \geq 0$, where $\I$ denotes the identity operator in $\H$, normalized by the Hilbert-space dimension. Since $W$ has unit trace and $\W$ is constructed to be a non-negative operator, the resulting AEW corresponds to a quantum state in $\H\otimes \H$. Note that the construction of AEWs from EWs is similar to structural physical approximation to positive maps, which is a transformation from positive maps to completely positive maps \cite{ref:spa}. The construction in Eq.~(\ref{eq:aewcon}) is analogous to the the structural physical approximation, see e.g. \cite{ref:spaconjecture}. 

The key idea in our scheme of entanglement detection is to apply AEWs instead of EWs, and to obtain the overlap $\tr[\rho \W]$ instead of $\tr[\rho W]$. For an AEW $\W$ from a witness $W$ with $p^{*}$, see Eq.~(\ref{eq:aewcon}), it holds that $0 \leq \tr[\rho \W] \leq 1$ for all quantum states $\rho \in S(\H \otimes \H)$. Also, the following relation between $W$ and $\W$ holds true for any state $\rho \in S(\H\otimes \H)$: $\tr[\rho \W] =(1-p^{*})\tr[\rho W] +p^{*}/d^2$ where $d$ denotes the dimension of Hilbert space $\H$. From this, we have the relation 
\bea
\tr[\rho W]  = \frac{1}{1-p^{*}} \left( \tr\left[\rho \W\right] - \frac{ p^{*}}{d^2}\right). \label{eq:relation}
\eea
where we recall that $p^{*}$ and $d$ are dictated by the construction. This shows that the overlap $\tr[\rho \W]$  allows us to learn expectation values of EWs, i.e., entanglement detection can be carried out. 

The overlap can be described as the fidelity of quantum states. With the fidelity $F(|\psi \rangle, |\phi\rangle) = |\langle \psi | \phi \rangle|^2$ for pure states, we introduce the average fidelity for mixed states $\rho = \sum_i p_i  | \psi_{i}^{(\rho)} \rangle \langle  \psi_{i}^{(\rho)} | $ and $\sigma= \sum_j q_j  | \phi_{j}^{(\sigma)} \rangle \langle  \phi_{j}^{(\sigma)}  |$ as,
\bea
F_{\mathrm{ave}}(\rho,\sigma) = \sum_{i,j} p_i q_j F (| \psi_{i}^{(\rho)} \rangle, | \phi_{j}^{(\sigma)} \rangle) \nonumber
\eea
which is equal to the overlap $\tr[\rho \sigma]$, and thus we have $F_{\mathrm{ave}}(\rho, \W) = \tr[\rho \W ]$. The average fidelity does not depend on a particular choice of decompositions of a given state. Note that if one of two arguments is a pure state, the overlap corresponds exactly to the Uhlmann fidelity \cite{ref:ufidelity}. Indeed, any experimental scheme capable of estimating the fidelity of two pure states can also be utilized to detect entangled states. By preparing a mixture of pure states in the scheme, one can estimate the average fidelity for a mixed state.  In the following, we are going to show that a HOM interferometer that generically estimates the fidelity of pure states of single-photon sources suffices for the detection of entangled states.

In general, AEWs need not be entangled. All results shown in the above with AEWs can also be extended to cases where they are separable. To be precise, in the construction in Eq.~(\ref{eq:aewcon}), minimal $p_s$ instead of $p^{*}$ can be chosen such that the resulting one $\W_s = (1-p_s)W +p_s \I$ is not only non-negative but also separable. Since the normalised identity $\I$ has full-rank, for any $W$ there always exists $p_s <1$ such that $\W_s$ is separable \cite{ref:fullrank}. In general, we have $p_{s}\geq p^{*}$ since the positivity condition of $\W$ does not necessarily imply its separability \cite{ref:kye, ref:darek, ref:tura}. It is known that $p_s = p^{*}$ for EWs which can detect all entangled isotropic states \cite{ref:darekiso, ref:fullrank}. In particular, let $\{ \W_s\}$ denote the set of separable AEWs (SAEWs). Then, in the same way, from the relation in Eq.~(\ref{eq:relation}) the detection scheme can in fact be performed by LOCC by preparing $\W_s$. 

For instance, consider a SAEW $\W_s$ that admits a separable decomposition as $\W_s = \sum_k p_k w_{k}^{(A)} \otimes w_{k}^{(B)}$ with quantum states $w_{k}^{(A)},$ $w_{k}^{(B)} \in S(\H)$. Entanglement detection for a distributed and unknown state $\rho_{AB}$ can be performed by estimating the average fidelity between the shared state and the separable state $\W_s$, i.e., $\tr[\rho_{AB}  \W_s]$, or also its fidelity with locally prepared product states $w_{k}^{(A)} \otimes w_{k}^{(B)}$, that is,
\bea
\tr\left[\rho_{AB} \W_s \right] & = & \sum_k p_k v_k,~ \nonumber \\
\mathrm{where} && v_k  =  \tr\left[ w_{k}^{(A)} \otimes w_{k}^{(B)} \rho_{AB}\right]. \label{eq:locc}
\eea
The two schemes of entanglement detection are operationally equivalent.

We now present an experimental proposal for detecting entangled states using a polarization-encoded photonic states with an HOM interferometry scheme. Since we need only realize the average fidelity $F_{\mathrm{ave}}(\W,\rho)$ in Eq.~(\ref{eq:relation}), any implementation that works towards this end can be generally applied. The goal here is to devise a feasible and efficient method, possibly with minimal resources in measurement. Our proposal  makes use of two techniques recently developed in quantum photonics, quantum joining and the embedding of orbital-angular-momentum degrees of freedom.

\begin{figure}[]
\includegraphics[width=3.0in,keepaspectratio]{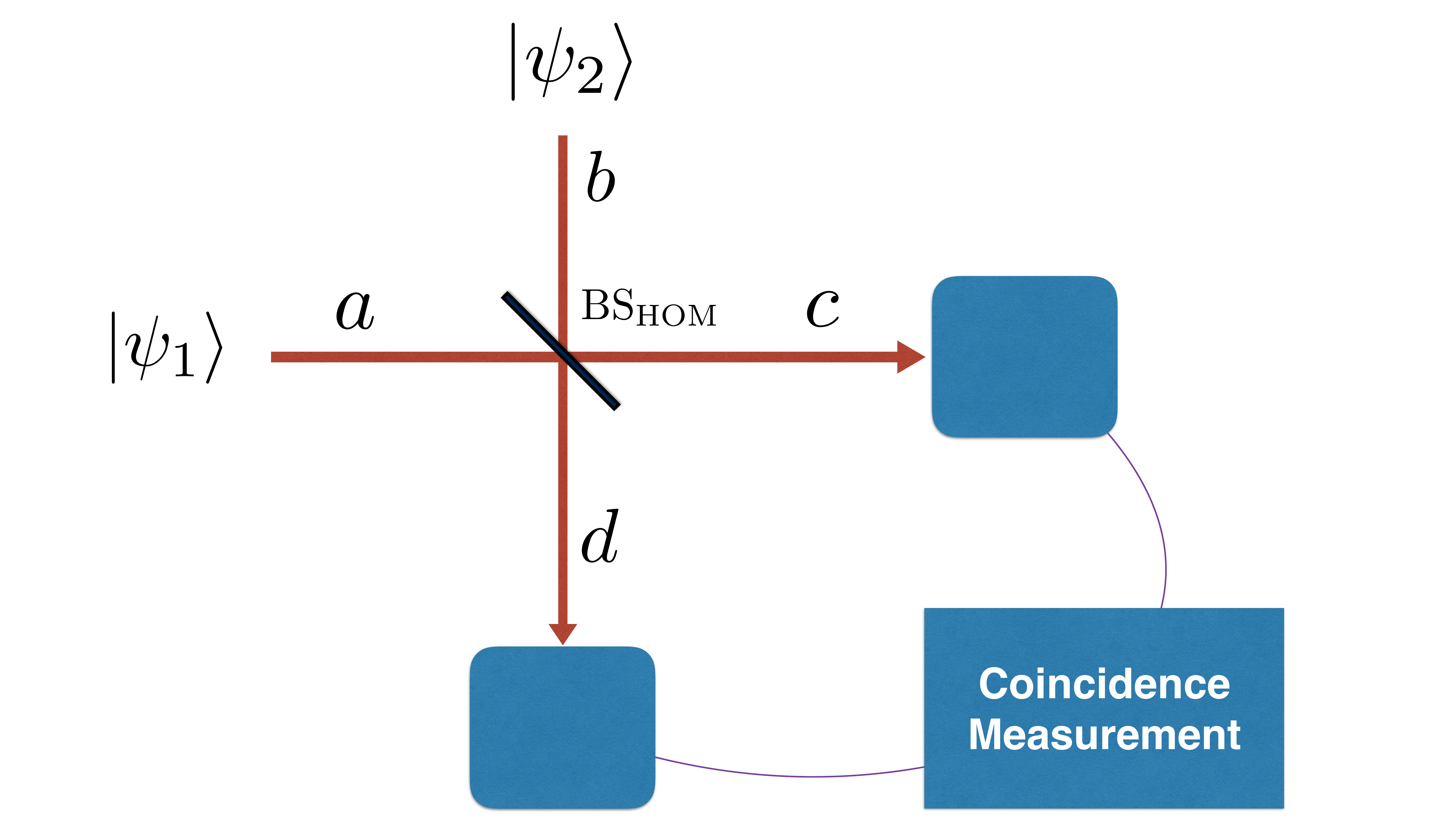}
\caption{The HOM interferometry is composed of a beam splitter and two detectors only. Each state $|\psi_i\rangle$ is denoted for a $d$-dimensional state of a single photon. The coincidence count $p_c$ is then measured by detectors at paths $c$ and $d$ and gives the fidelity $F (|\psi_1\rangle, |\psi_2 \rangle) = |\langle \psi_1 | \psi_2 \rangle |^2$, see Eq.~(\ref{eq:esti}). }
\label{fig:HOM}
\end{figure}

Let $\W$ denote a known state derived from witness $W$, and  $\rho$ is a given unknown state. One finds a decomposition of $\W$, and there also exists a decomposition of state $\rho$, as follows:
\bea
\widetilde{W} = \sum_i \alpha_i  | \phi_i\rangle \langle \phi_i |,~~\mathrm{and}~ \rho = \sum_j \beta_j | \Psi_j\rangle \langle \Psi_j |.  \label{eq:decom}
\eea
with states $\{ |\phi_i \rangle \}$ that are not necessarily orthogonal, and unknown states $\{ |\Psi_j  \rangle \}$. From the relation
\bea
\mathrm{tr}\left[ \rho \W \right] = \sum_{i,j} \alpha_i \beta_j  | \langle \phi_i |  \Psi_j \rangle |^2, \label{eq:esti}
\eea
the overlap, that is, the average fidelity $F_{\mathrm{ave}} (\W,\rho)$ in the left-hand-side can be obtained if an experimental scheme estimates the fidelity $F(|\phi_i\rangle, |\Psi_j \rangle)$ in the right-hand side. 
Since the experiment for detection of entanglement of pure state can be extended straightforwardly to mixed state by generalising the state preparation from pure to mixed state, it is sufficient to demonstrate our experiment proposal for the case of pure states. Hence, without any loss of generality, we focus only on the case of pure states. In what follows, we detail the proposal for detecting a two-qubit entangled state.

\begin{figure}[]
\includegraphics[width=3.5in,keepaspectratio]{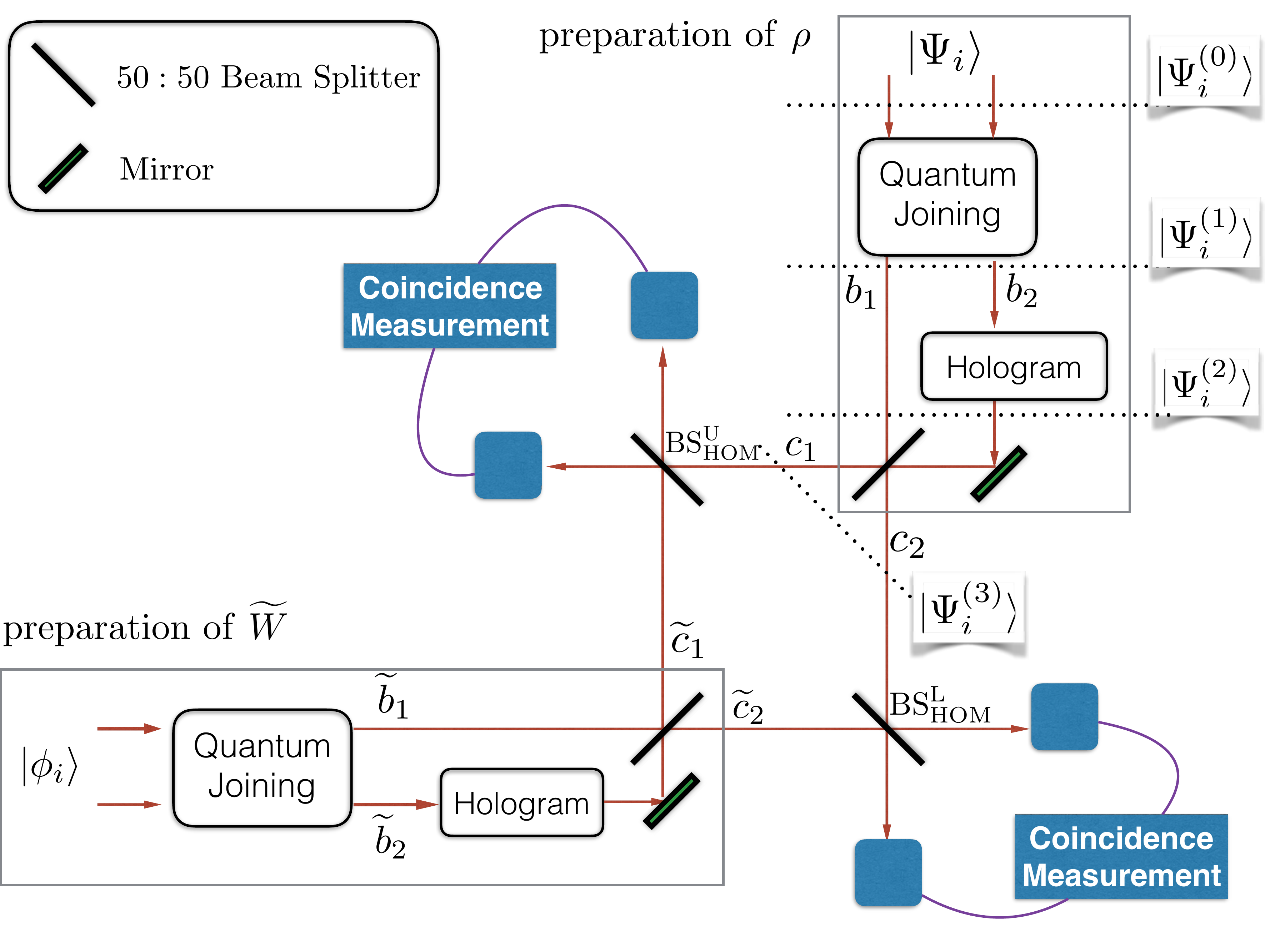}
\caption{The implementation scheme to estimate $\tr[\W \rho]$ is shown. A two-photon polarisation state $\rho$ is initially prepared as a mixture of $\{\beta_i ,|\Psi_i\rangle \}$, so is $\W$ with $\{\alpha_i, |\phi_i\rangle \}$, see Eq.~(\ref{eq:decom}). After quantum joining, a single-photon polarisation is produced with paths $b_1$ or $b_2$, $|\Psi^{(1)}\rangle$ and, depending on the path, OAM degrees of freedom is encoded by the hologram, $|\Psi^{(2)}\rangle$. Two spatial modes interfere via the $50:50$ beam splitter: the resulting single-photon four-dimensional state $|\Psi^{(3)}\rangle $ of polarisation and OAM degrees of freedom appears in $c_1$ or $c_2$ with probabilities $1/2$, respectively. The same procedure applies to state $\W$, which is found in $\widetilde{c}_1$ or $\widetilde{c}_2$ with probabilities $1/2$, respectively. The HOM effect appears when two photons pass the same beam splitter together, which happens with the overall probability $1/2$. }
\label{fig:proposal}
\end{figure}

The experimental proposal for entanglement detection is shown in Fig.~\ref{fig:proposal}.  Depending on the required efficiency, the set-up can estimate $\tr[\rho\W]$ in Eq.~(\ref{eq:relation}) with just two or four detectors regardless of dimensions and the number of modes of given states. We fully exploit the HOM interferometry for estimating the visibility of the two single-photon input states with coincidence measurement. The measured visibility in fact corresponds to the average fidelity of single-photon states. If two single photons enter two input ports of a 50 : 50 beamsplitter, and detectors are placed at the output ports, then the probability of detecting a simulatneous events in both ports of the interferometer, namely the coincidence count rate, is given by $p_c$. For states $\sigma_1$ and $\sigma_2$ of single photons, the coincidence probability is given as, $p_c = (1-\tr[\sigma_1 \sigma_2])/2$, from which
\bea
\tr\left[\sigma_1 \sigma_2\right] =  1 - 2 p_c. \label{eq:pc}
\eea
By observing the coincidence count rate $p_c$, the overlap of single-photon states can be found. We recall that the HOM effect holds for high-dimensional states of a single photon \cite{ref:GarciaEscartin}. Degrees of freedom of photons can be the optical frequency, OAM states \cite{ref:Slussarenko, ref:Nagali, ref:Nagali2} or time-bin encoding \cite{ref:Donohue, ref:Yu}. It is worth  noting that the measurement setting of the HOM effect involves only two detectors, see Fig.~\ref{fig:HOM}.

Let us now explain the detection scheme shown in Fig.~\ref{fig:proposal} in detail. The preparation corresponds to a mixture of $\{\beta_i ,|\Psi_i\rangle \}$ for state $\rho$ and $\{\alpha_i, |\phi_i\rangle \}$ for state $\W$ of polarisation degrees of freedom of two photons, respectively. To realize quantum state joining and the embedding of OAM degrees of freedom, let us begin with a two-photon polarization state, say $|\Psi^{(0)}\rangle$, where horizontal and vertical polarisations are denoted by $|H\rangle$ and $|V\rangle$, respectively,
\bea
|\Psi^{(0)} \rangle = x_0 \Ket{H, H}+ x_1 \Ket{H, V} + x_2 \Ket{V, H} + x_3 \Ket{V, V}. \nonumber
\eea
Note that the quantum state is fully characterized by coefficients $(x_0,x_1,x_2,x_3)$. The following procedures for the state in the above apply to both states $\rho$ and $\W$.

We first apply quantum joining which does not change the quantum state, and yet the polarization degrees of freedom of two photons are converted into a single photon's polarization and spatial degrees of freedom \cite{ref:quantumjoin}, see also the general theoretical framework \cite{ref:passaro}. The quantum joining process works probabilistically and is described in Supplementary Material in detail. The resulting polarization state can be found in either $b_1$ or $b_2$ mode, see also Fig.~\ref{fig:proposal}
\bea
| \Psi^{(1)} \rangle = x_0 \Ket{H_\mathrm{b_1}} + x_1 \Ket{V_ \mathrm{b_1}} + x_2 \Ket{H_\mathrm{b_2}} + x_3 \Ket{V_\mathrm{b_2}}, \nonumber
\eea
where $|A_a\rangle$ denotes a photon's polarisation $A$ at $a$ spatial mode.  


In order to be able to apply the HOM interferometer at the end, the spatial mode of the photon is mapped to a static degree of freedom of the photon, i.e., each input port of the beam splitters must consist of only one single photon. We exploit OAM degrees of freedom of a single photon \cite{ref:eOAM} to erase the information in the spatial mode. We place a hologram in path $\mathrm{b}_2$ that introduces a non-zero OAM, say $q$, for a photon in the path, and obtain the state
\bea
| \Psi^{(2)}  \rangle &=& x_0 \Ket{H_\mathrm{b_1},0} + x_1 \Ket{V_\mathrm{b_1},0}+ x_2 \Ket{H_\mathrm{b_2},q } + x_3 \Ket{V_\mathrm{b_2},q}.~~~ \nonumber
\eea
To erase the spatial mode information of the photon, we place a $50:50$ beam splitter that combines paths $b_1$ and $b_2$ to have
\bea
| \Psi^{(3)}  \rangle &=& x_0 \Ket{H ,0} + x_1 \Ket{V ,0}+ x_2 \Ket{H , q } + x_3 \Ket{V,q}. \nonumber
\eea
The state $| \Psi^{(3)}  \rangle$ after the $50:50$ beam splitter appears in path $c_1$ and $c_2$ with probabilities $1/2$, respectively. Throughout, the quantum state characterized by $(x_0,x_1,x_2,x_3)$ remains unchanged. 

The same procedure applies to the preparation of state $\W$, i.e. quantum joining, OAM encoding, and collecting to a single path. After all, the resulting single-photon state $\W$ is found at $\widetilde{c}_1$ or $\widetilde{c}_2$ with probabilities $1/2$, respectively. Together with the other single-photon source of state $\rho$ from mode $c_1$ or $c_2$, two single photons of $\W$ and $\rho$ interfere with each other at beam splitters $\mathrm{BS_{HOM}^{U}}$ or $\mathrm{BS_{HOM}^{L}}$, each of which happens with probability $1/4$, respectively.



To detect entanglement, we suppose that copies of states $\W$ and $\rho$ are initially prepared and the experimental set-up shown in Fig.~\ref{fig:proposal} is applied. The goal is to detect entanglement of the unknown state $\rho$. Once $N$ copies of them, $\W^{\otimes N}$ and $\rho^{\otimes N}$, undergoes quantum state joining, the cases where they interfere each other via the beam splitters, $\mathrm{BS_{HOM}^{U}}$ or $\mathrm{BS_{HOM}^{L}}$, occurs when photons are found in $( c_1, \widetilde{c}_1)$ or $(c_2 ,\widetilde{c}_2)$. The other cases corresponding to photons  along paths $(c_1,\widetilde{c}_2)$ or $(\widetilde{c}_1, c_2)$ do not interfere  and they are discarded. Note that $50:50$ beam splitters are applied when two paths are combined for the OAM encoding, hence only $N/2$ copies contributed to the HOM effect. Let $N_{c}^{U}$ and $N_{c}^{L}$ denote the number of coincidence detection events at $\mathrm{BS_{HOM}^{U}}$ or $\mathrm{BS_{HOM}^{L}}$ respectively. Note that we have $N_{c}^{U} = N_{c}^{L} := N_c$ due to $50:50$ beam splitters. Then, we have from Eq.~(\ref{eq:pc}),
\bea
1-2 p_c = 1- 2\times \frac{N_{c}^{U} + N_{c}^{L }}{N\times \frac{1}{2}} = 1- 2\times \frac{2 N_c}{N\times \frac{1}{2}}. \nonumber
\eea
From Eq.~(\ref{eq:relation}) and Eq.~(\ref{eq:pc}), we have obtained, straightforwardly, for two-qubit states ($d=2$),
\bea
\tr[\rho W] = \frac{1}{1-p^{*}} \left[ 1- \frac{8N_c}{N} - \frac{p^{*}}{4} \right], \nonumber 
\eea
with the observed quantity $p_c = 4N_c / N$. Note that $p^{*}$ is given in Eq.~(\ref{eq:aewcon}) such that $\W\geq0$, depending on a witness $W$. In the above, we emphasize that $p_c$ is the only quantity targetted in the experiment. Therefore, a given state $\rho$ is entangled if $\tr[\rho W]<0$, i.e., the observed coincidence count rate in experiment satisfies that
\bea
N_c > \frac{N}{8} ( 1- \frac{p^{*}}{4}),~~\mathrm{for~given}~ N~ \mathrm{and}~p^{*}.  \nonumber
\eea
We stress that we need only a single HOM interferometer with two detectors to find the expectation $\tr[\rho W]$ experimentally. The optical paths labeled $c_1$ and $\tilde{c}_1$ in Fig. \ref{fig:proposal} could be dumped. This would simplify the implementation complexity, at the cost of reducing the probability of success by a 1/2 factor, i.e. only $N/4$ copies would contribute to the coincidence count.

In conclusion, we have shown that any scheme estimating the average fidelity of two quantum states can be applied to detecting entangled states, and also that all entangled states can be detected via this scheme. In particular, we present an experimental proposal with a single HOM interferometer in the measurement where only two detectors are applied regardless of the system dimension. This contrasts vividly with the case of quantum tomography where the number of detectors increases with dimensions and modes of a given system. The detection scheme can be in principle generalized to high-dimensional and multipartite systems, with the current measurement set-up but, incorporating more OAM degrees of freedom. From a fundamental perspective, our work shows that the detection of entanglement demands significantly less information compared to a full tomographical. In short, experimental resources of EWs are generally inequivalent to those of state tomography.

Our proposal makes use of quantum state joining and the embedding of OAM degrees of freedom prior to any measurement. The original scheme in Ref. \cite{ref:quantumjoin} applies post-selection, hence probabilistic, which comes from the fact that photon sources are applied. In principle, quantum state joining can be performed generally in a trace-preserving manner \cite{ref:passaro}. In addition, the application of other degrees of freedom such as time-bin encoding can be used to replace the OAM degrees of freedom, i.e. the implementation of our scheme is not limited to OAM en- coded states.

We remark that all results here hold for cases of SAEWs detecting entangled states through LOCC, and this technique using SAEWs can be highly useful for distributed entangled states. We also remark that the experimental proposal can be extended to higher-dimensional  multipartite quantum states, that is, EWs in $\{W^{\dagger} = W \in \B(\H\otimes \H\otimes \cdots \otimes \H)\}$. For this purpose, in the experimental part shown in Fig.~\ref{fig:proposal}, the preparation stage involves more steps and more resources of OAM. Quantum joining can also be generalized to higher-dimensional states, see e.g.  \cite{ref:passaro}, exploting more OAM degrees of freedom. Nonetheless, the measurement set-up rem,ains the same with a single HOM interferometer and two detectors. In the multipartite cases, it would be interesting to develop a detection scheme to cover a finer structure of convex sets of quantum states, e.g. detection of $k$-separable states or genuinely entangled states, etc. depending on the properties of EWs. 

Finally,  we reiterate that our scheme have provided a way of estimating expectation values of EWs in an efficient and feasible way. In fact, the expectation values of EWs have various applications, not just for entanglement detection but also for the estimation of the fidelity of multipartite states in which the equivalent resources for tomography are highly non-trivial, see e.g. \cite{ref:toth}. We envisage that our proposal could be useful for quantum information applications that exploit expectation values of Hermitian operators in general. 

\subsection*{Acknowledgment}

C.J.K. and L.C.K. acknowledge support from the National Research Foundation \& Ministry of Education, Singapore. S.F. acknowledges the financial support from UPV/EHU UFI 11/55 and University Sorbonne Paris Cité EQDOL contract. J.B. is supported by Institute for Information \& communications Technology Promotion(IITP) grant funded by the Korea government(MSIP) (No.R0190-15-2028, PSQKD) and the KIST Institutional Program (Project No. 2E26680-16-P025).


\newpage 

\section{\textit{Supplemental Material:} Entanglement Detection with Single Hong-Ou-Mandel Interferometry}

\subsection{The Hong-Ou-Mandel interferometry for four-dimensional states}

In the paper, we in particular consider the Hong-Ou-Mandel (HOM) effect for four-dimensional single-photon states. Here, we discuss our experimental proposal in details. In order to implement the multidimensional HOM interferometer required for entanglement detection, we merge the state encoded in a polarization of different photons into the orbital angular momentum (OAM) state of a single photon. Note that the HOM effect is not limited to qubit states but also appears in $d$-dimensional quantum states in general. Lets consider two 4-dimensional quantum states, $\Ket{\psi_1}=\sum^3_{i=0} a_i \Ket{i}$ and $\Ket{\psi_2}=\sum^3_{j=0} b_i \Ket{j}$. Upon passing through a beam splitter (see Fig. \ref{fig:HOM}), we have
\begin{eqnarray}
\begin{aligned}
	\hat{a}^\dag_i &\xrightarrow{BS}& \frac{1}{\sqrt{2}}\left(\hat{c}^\dag_i+\hat{d}^\dag_i \right)\\
	\hat{b}^\dag_i &\xrightarrow{BS}& \frac{1}{\sqrt{2}}\left(\hat{c}^\dag_i-\hat{d}^\dag_i \right).
	\end{aligned}
	\label{eqn:bs}
\end{eqnarray}
where $\hat{a}^\dag_i (\hat{b}^\dag_j)$ creates a photon with state $\Ket{i} (\Ket{j})$ at mode $a (b)$. \\
Hence, from Eqn. (\ref{eqn:bs}), we obtain
\begin{widetext}
\begin{eqnarray*}
	\Ket{\psi_1}\Ket{\psi_2} &=& \left( \sum^3_{i=0} a_i \hat{a}^\dag_i\right) \left( \sum^3_{j=0} b_j \hat{b}^\dag_j\right) \Ket{vac},\\
	\Ket{\psi_1}\Ket{\psi_2} &\xrightarrow{BS}& \frac{1}{2} \left[ \sum^3_{i=0} a_i \left(\hat{c}^\dag_i+\hat{d}^\dag_i \right) \right] \left[ \sum^3_{j=0} b_j \left(\hat{c}^\dag_j-\hat{d}^\dag_j \right) \right] \Ket{vac}\\
	&=& \frac{1}{2} \sum_i \sum_j \left[ a_i b_j \left( c^\dag_i c^\dag_j - d^\dag_i d^\dag_j \right) + a_i b_j \left( -c^\dag_i d^\dag_j + c^\dag_j d^\dag_i \right) \right]\Ket{vac}.
\end{eqnarray*}
\end{widetext}
where $\Ket{vac}$ represents the vacuum state.

\begin{figure}[ht!]
\includegraphics[width=3.0in,keepaspectratio]{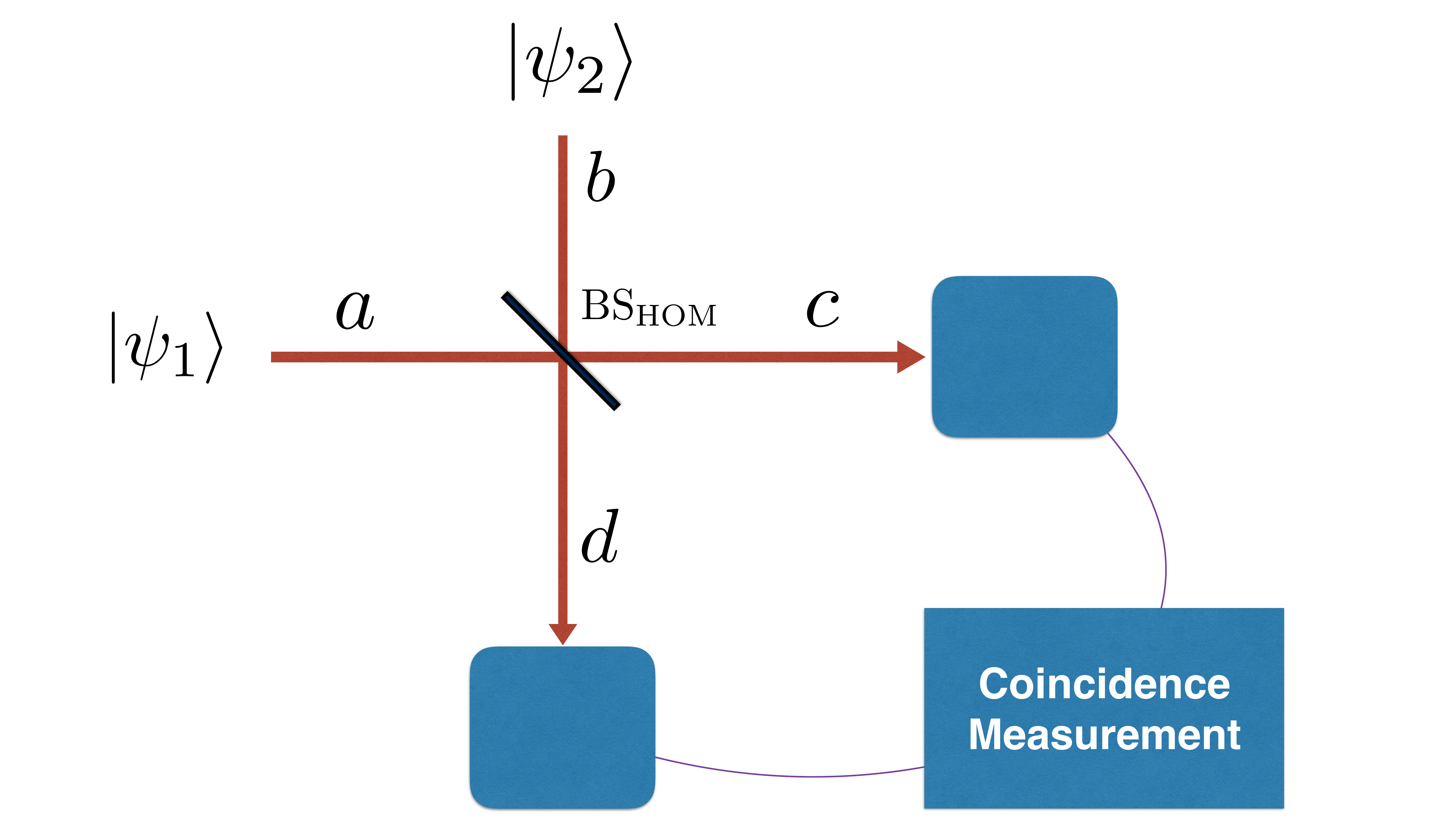}
\caption{The HOM interferometry is composed of a beam splitter and two detectors only. Each state $|\psi_i\rangle$ is denoted for a $d$-dimensional state of a single photon. The coincidence count $p_c$ is then measured by detectors at paths $c$ and $d$ and gives the visibility $|\langle \psi_1 | \psi_2 \rangle |^2$. }
\label{fig:HOM}
\end{figure}

For the coincidence detection of photons at path $c$ and $d$, we have
\begin{eqnarray*}
&& \frac{1}{2} \sum^3_{i=0} \sum^3_{j=0}\left[a_i b_j \left( -c^\dag_i d^\dag_j + c^\dag_j d^\dag_i \right)\right] \nonumber \\
&&	= \frac{1}{2} \sum^3_{i=0} \sum^3_{j=0}\left[\left(a_j b_i-a_i b_j\right) c^\dag_i d^\dag_j \right],
\end{eqnarray*}
by swapping the indices on the second term of the coincidence count. As a result, the probability of coincidence count, $p_c$ is given by
\begin{eqnarray*}
	p_c&=& \left| \frac{1}{2} \left( a_j b_i - a_i b_j \right) \right|^2\\
	 &=& \sum_i \sum_j \frac{1}{4}  \left(a^*_j b^*_i -a^*_i b^*_j \right)\left(a_j b_i-a_i b_j\right)\\
	&=& \sum_i \sum_j \frac{1}{4} \left( 2\left|a_i\right|^2 \left|b_j\right|^2 - a_i a^*_j b^*_i b_j - a_j a^*_i b^*_j b_i \right)\\
	&=& \sum_i \sum_j \frac{1}{2} \left( \left|a_i\right|^2 \left|b_j\right|^2 - a_i a^*_j b^*_i b_j \right)\\
	&=& \frac{1}{2} \left( 1 - \left| \Braket{\psi_2|\psi_1} \right|^2\right),
\end{eqnarray*}
where $\left| \Braket{\psi_2|\psi_1} \right|^2 = \sum_i \sum_j a_i a^*_j b_j b^*_i$.


\subsection{Quantum Joining Setup}

\begin{figure}[ht!]
\includegraphics[width=3.4in,keepaspectratio]{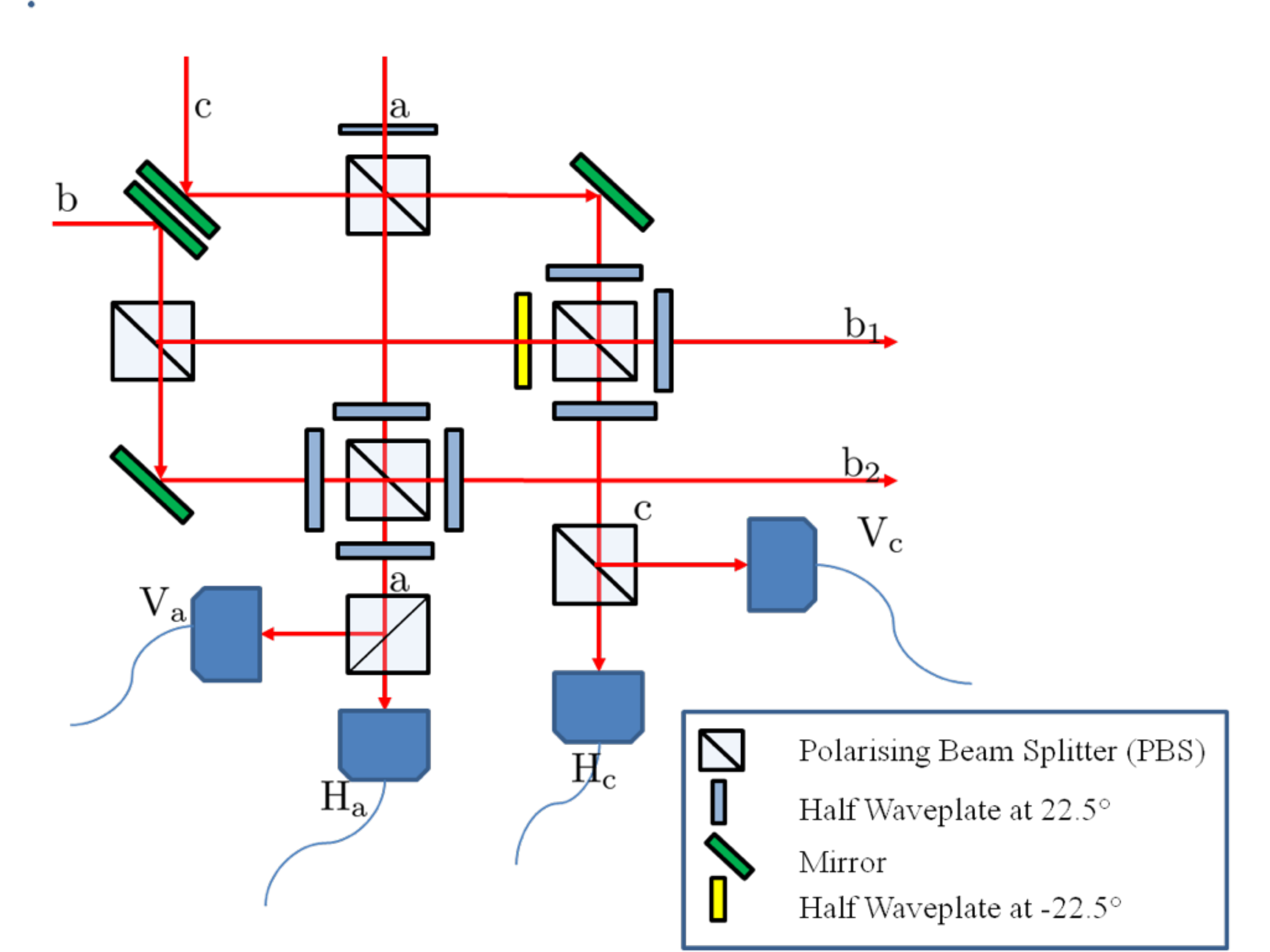}
\caption{ (Colour online) The experimental setup for the quantum joining of two polarization encoded photonic qubits into a single photonic qubit which is polarization and path encoded. The two input states are injected at path c and path b while an ancillary photon at path a. All three states are encoded in linear polarization states with the ancilla photon is horizontally polarised while the two input photons are in arbitrary superposition of horizontally and vertically polarised states. The half waveplate are oriented at $22.5\degree$ with the exception of the half waveplate coloured yellow that is oriented at $-22.5\degree$. Upon post-selection of coincidence count of detectors $\mathrm{H_a}$ and $\mathrm{H_c}$, we obtain a four dimensional output state (at path $\mathrm{b_1}$ and $\mathrm{b_2}$) encoded in polarization and path. }
\label{fig:JoiningSetup}
\end{figure}

The quantum joining of an arbitrary two photon states encoded in polarization into a four dimensional single photon state can be done experimentally as shown in Ref. \cite{qJoin}. The experimental setup is shown in Fig. \ref{fig:JoiningSetup}. 

Considering a generic initial state, $\Ket{\psi}=\left( \alpha_0 H_b H_c + \alpha_1 H_b V_c + \alpha_2 V_b H_c + \alpha_3 V_b V_c \right) \mathrm{H_a}$, we can trace the evolution of the state upon passing through the different optical components in Fig. \ref{fig:JoiningSetup}. Let $HWP_{x,k}$ denotes the half-wave plate placed at path $x$ while $k$ represents the labelling of half-wave plates in ascending order as counted from the input port of the experimental setup. On the other hand, $PBS_{y,z}$ denotes the polarising beam splitter placed such that it mixes the path $y$ and $z$. We will then have
\begin{widetext}
\begin{eqnarray*}
\Ket{\psi} &\xrightarrow{HWP_{a,0}}& \frac{1}{\sqrt{2}} \left( \alpha_0 H_b H_c + \alpha_1 H_b V_c + \alpha_2 V_b H_c + \alpha_3 V_b V_c \right) \left( H_a + V_a \right)\\
&\xrightarrow{PBS_{a,c}}& \frac{1}{\sqrt{2}} \left( \alpha_0 H_b H_c + \alpha_1 H_b V_a + \alpha_2 V_b H_c + \alpha_3 V_b V_a \right) \left( H_a + V_c \right)\\
&\xrightarrow{PBS_{b_1,b_2}}& \frac{1}{\sqrt{2}} \left( \alpha_0 H_{b_1} H_c + \alpha_1 H_{b_1} V_a + \alpha_2 V_{b_2} H_c + \alpha_3 V_{b_2} V_a \right) \left( H_a + V_c \right)\\
&\xrightarrow{HWP_{a,1}}& \frac{1}{2} \left[ \alpha_0 H_{b_1} H_c + \alpha_1 H_{b_1} \left( H_a - V_a \right) + \alpha_2 V_{b_2} H_c + \alpha_3 V_{b_2} \left( H_a - V_a \right) \right] \left( H_a + V_a + V_c \right)\\
&\xrightarrow{HWP_{c,1}}& \frac{1}{2 \sqrt{2}} \left[ \alpha_0 H_{b_1} \left( H_c+ V_c \right) + \alpha_1 H_{b_1} \left( H_a - V_a \right) + \alpha_2 V_{b_2} \left( H_c+ V_c \right) + \alpha_3 V_{b_2} \left( H_a - V_a \right) \right] \left( H_a + V_a + H_c - V_c  \right)\\
&\xrightarrow{HWP_{b_1,1}}& \frac{1}{4} \left[ \alpha_0 \left( H_{b_1} + V_{b_1}\right) \left( H_c+ V_c \right) + \alpha_1 \left( H_{b_1} + V_{b_1}\right) \left( H_a - V_a \right) + \alpha_2 V_{b_2} \left( H_c+ V_c \right) + \alpha_3 V_{b_2} \left( H_a - V_a \right) \right] \left( H_a + V_a + H_c - V_c  \right)\\
&\xrightarrow{HWP_{b_2,1}}& \frac{1}{4} \left[ \alpha_0 \left( H_{b_1} + V_{b_1}\right) \left( H_c+ V_c \right) + \alpha_1 \left( H_{b_1} + V_{b_1}\right) \left( H_a - V_a \right) + \alpha_2 \left( H_{b_2} + V_{b_2}\right) \left( H_c+ V_c \right) \right.\\
&& \left.+ \alpha_3 \left( H_{b_2} + V_{b_2}\right) \left( H_a - V_a \right) \right] \left( H_a + V_a + H_c - V_c  \right)\\
&\xrightarrow{PBS_{a,b_1}}& \frac{1}{4} \left[ \alpha_0 \left( H_{b_1} + V_{a}\right) \left( H_c+ V_c \right) + \alpha_1 \left( H_{b_1} + V_{a}\right) \left( H_a - V_{b_1} \right) + \alpha_2 \left( H_{b_2} + V_{b_2}\right) \left( H_c+ V_c \right) \right. \\
&& \left.+ \alpha_3 \left( H_{b_2} + V_{b_2}\right) \left( H_a - V_{b_1} \right) \right] \left( H_a + V_{b_1} + H_c - V_c  \right)\\
&\xrightarrow{PBS_{c,b_2}}& \frac{1}{4} \left[ \alpha_0 \left( H_{b_1} + V_{a}\right) \left( H_c+ V_{b_2} \right) + \alpha_1 \left( H_{b_1} + V_{a}\right) \left( H_a - V_{b_1} \right) + \alpha_2 \left( H_{b_2} + V_{c}\right) \left( H_c+ V_{b_2} \right) \right. \\
&& \left.+ \alpha_3 \left( H_{b_2} + V_{c}\right) \left( H_a - V_{b_1} \right) \right] \left( H_a + V_{b_1} + H_c - V_{b_2}  \right) \\
&\xrightarrow{HWP_{a,2}}& \frac{1}{4} \left[ \alpha_0 \left( H_{b_1} + \frac{1}{\sqrt{2}}\left( H_a - V_a \right)\right) \left( H_c+ V_{b_2} \right) + \alpha_1 \left( H_{b_1} + \frac{1}{\sqrt{2}}\left( H_a - V_a \right)\right) \left( \frac{1}{\sqrt{2}}\left( H_a + V_a \right) - V_{b_1} \right) \right. \\
&& \left. + \alpha_2 \left( H_{b_2} + V_{c}\right) \left( H_c+ V_{b_2} \right) + \alpha_3 \left( H_{b_2} + V_{c}\right) \left( \frac{1}{\sqrt{2}}\left( H_a + V_a \right) - V_{b_1} \right) \right] \left( \frac{1}{\sqrt{2}}\left( H_a + V_a \right) + V_{b_1} + H_c - V_{b_2}  \right)\\
&\xrightarrow{HWP_{c,2}}& \frac{1}{4} \left[ \alpha_0 \left( H_{b_1} + \frac{1}{\sqrt{2}}\left( H_a - V_a \right)\right) \left( \frac{1}{\sqrt{2}}\left( H_c + V_c \right)+ V_{b_2} \right) + \alpha_1 \left( H_{b_1} + \frac{1}{\sqrt{2}}\left( H_a - V_a \right)\right) \left( \frac{1}{\sqrt{2}}\left( H_a + V_a \right) - V_{b_1} \right) \right. \\
&& \left. + \alpha_2 \left( H_{b_2} + \frac{1}{\sqrt{2}} \left( H_c - V_c \right)\right) \left( \frac{1}{\sqrt{2}} \left( H_c + V_c \right)+ V_{b_2} \right) + \alpha_3 \left( H_{b_2} + \frac{1}{\sqrt{2}} \left( H_c - V_c \right)\right) \left( \frac{1}{\sqrt{2}}\left( H_a + V_a \right) - V_{b_1} \right) \right]  \\
&& \left( \frac{1}{\sqrt{2}}\left( H_a + V_a \right) + V_{b_1} + \frac{1}{\sqrt{2}} \left( H_c + V_c \right) - V_{b_2}  \right)\\
&\xrightarrow{HWP_{b_1,2}}& \frac{1}{4} \left[ \alpha_0 \left( \frac{1}{\sqrt{2}} \left( H_{b_1} + V_{b_1} \right) + \frac{1}{\sqrt{2}}\left( H_a - V_a \right)\right) \left( \frac{1}{\sqrt{2}}\left( H_c + V_c \right)+ V_{b_2} \right) \right.\\
&& \left. + \alpha_1 \left( \frac{1}{\sqrt{2}} \left( H_{b_1} + V_{b_1} \right) + \frac{1}{\sqrt{2}}\left( H_a - V_a \right)\right) \left( \frac{1}{\sqrt{2}}\left( H_a + V_a \right) - \frac{1}{\sqrt{2}} \left( H_{b_1} - V_{b_1} \right) \right) \right. \\
&& \left. + \alpha_2 \left( H_{b_2} + \frac{1}{\sqrt{2}} \left( H_c - V_c \right)\right) \left( \frac{1}{\sqrt{2}} \left( H_c + V_c \right)+ V_{b_2} \right) \right. \\
&& \left. + \alpha_3 \left( H_{b_2} + \frac{1}{\sqrt{2}} \left( H_c - V_c \right)\right) \left( \frac{1}{\sqrt{2}}\left( H_a + V_a \right) - \frac{1}{\sqrt{2}} \left( H_{b_1} - V_{b_1} \right) \right) \right]  \\
&& \left( \frac{1}{\sqrt{2}}\left( H_a + V_a \right) + \frac{1}{\sqrt{2}} \left( H_{b_1} - V_{b_1} \right) + \frac{1}{\sqrt{2}} \left( H_c + V_c \right) - V_{b_2}  \right)\\
&\xrightarrow{HWP_{b_2,2}}& \frac{1}{8\sqrt{2}} \left[ \alpha_0 \left( H_{b_1} + V_{b_1} + H_a - V_a \right) \left( H_c + V_c + H_{b_2} - V_{b_2} \right) + \alpha_1 \left( H_{b_1} + V_{b_1}  +  H_a - V_a \right) \left( H_a + V_a - H_{b_1} + V_{b_1}  \right) \right. \\
&& \left. + \alpha_2 \left( H_{b_2} + V_{b_2} + H_c - V_c \right) \left( H_c + V_c +  H_{b_2} - V_{b_2} \right) \right. 
\nonumber
\end{eqnarray*}
\end{widetext}
\begin{widetext}
\begin{eqnarray*}
&& \left. + \alpha_3 \left(  H_{b_2} + V_{b_2}  + H_c - V_c \right) \left( H_a + V_a -  H_{b_1} + V_{b_1}  \right) \right] \left( H_a + V_a + H_{b_1} - V_{b_1} + H_c + V_c - H_{b_2} + V_{b_2}  \right).
\end{eqnarray*}
Upon post-selecting the states with a photon detected in path $a$ and $c$ each, we have
\begin{eqnarray*}
\Ket{\psi} &=& \frac{1}{4\sqrt{2}} \left[ H_a H_c \left( \alpha_0 H_{b_1} +\alpha_1 V_{b_1} + \alpha_2 H_{b_2} + \alpha_3 V_{b_2} \right) \right.\\
&&+ \hspace{0.5cm} H_a V_c \left( \alpha_0 H_{b_1} +\alpha_1 V_{b_1} + \alpha_2 V_{b_2} + \alpha_3 H_{b_2} \right) \\
&& + \hspace{0.5cm} V_a H_c \left( \alpha_0 V_{b_1} +\alpha_1 H_{b_1} + \alpha_2 H_{b_2} + \alpha_3 V_{b_2} \right) \\
&&  \left. + \hspace{0.5cm} V_a V_c \left( \alpha_0 V_{b_1} +\alpha_1 H_{b_1} + \alpha_2 V_{b_2} + \alpha_3 H_{b_2} \right) \right].
\end{eqnarray*}
\end{widetext}

In order to obtain the state required in our scheme, we post-select the state when the photon in path $a$ and $c$ are both horizontally polarised. The post-selection is done with a success probability of 1/32.

\subsection{OAM encoding}
In the previous section, we have reviewed in details how a two-qubit state, encoded in the polarization of two photons, can be joined into the state of a single photon using polarization and spatial degrees of freedom. In order to implement the protocol proposed in the main text, we need to transform the information about the path into OAM encoding. This can be done by transforming the OAM state of photons propagating in one of the two path, and then merging the two paths into one in a beam splitter. The last step entails a 50\% loss of the overall signal.

The most compelling tools to modify the OAM state of light in the quantum regime are spatial light modulators~\cite{bazhenov1990laser,heckenberg1992generation,leach2002measuring} (SLM) and q-plates~\cite{marrucci2006optical,nagali2009quantum,marrucci2011spin}. SLMs make use of computer-generated holograms that works in a similar way to diffraction gratings. They are flexible devices as they work with a broad frequency range and they can be insensitive to polarization state~\cite{liu2015demonstration}. Q-plates are non-uniform birefringent plates that induce a spin-orbit coupling between the polarization and OAM degrees of freedom, at the single-photon level. Accordingly, the effect of q-plates corresponds to an entangling gate in the polarization and OAM state. The SLM are more suitable for our purposes, as we require to apply a fixed amount of OAM to photons crossing a given path, independently of their polarization state.

\end{document}